\documentclass[useAMS,usenatbib]{mnras}

\usepackage{graphicx}
\usepackage{natbib}
\usepackage{amssymb}
\usepackage{amsmath}
\usepackage{setspace}
\usepackage{epsfig,color}
\usepackage{multicol}
\usepackage{hyperref}
\usepackage{upgreek} 
\usepackage{longtable}
\usepackage[normalem]{ulem}

\newcommand{\modify}[1]{\textcolor{blue}{#1}}

\long\def\symbolfootnote[#1]#2{\begingroup
\def\thefootnote{\fnsymbol{footnote}}\footnote[#1]{##https://www.overleaf.com/project/5b98c85301ba633150bd62a42}\endgroup} 

\title{Density profile of a self-gravitating polytropic turbulent fluid in a rotating disk near to the cloud core}
\title[Density profile of a rotating fluid]{Density profile of a self-gravitating polytropic turbulent fluid in a rotating disk near to the cloud core}

\author[Donkov et al.]
{
\parbox{\textwidth}{S. Donkov$^{1\,\star}$, I. Zh. Stefanov$^2$, T. V. Veltchev$^{3, 4}$ and R. S. Klessen$^{4, 5}$}\vspace{0.4cm} \\
\parbox{\textwidth}{
  $^1$Institute of Astronomy and NAO, Bulgarian Academy of Sciences, Blvd. Tsarigradsko Shosse 72, 1000 Sofia, Bulgaria \\
  $^2$Department of Applied Physics, Technical University, 8 Kliment Ohridski Blvd., 1000 Sofia, Bulgaria \\
  $^3$University of Sofia, Faculty of Physics, 5 James Bourchier Blvd., 1164 Sofia, Bulgaria \\
  $^4$Universit\"at Heidelberg, Zentrum f\"ur Astronomie, Institut f\"ur Theoretische Astrophysik, Albert-Ueberle-Str. 2, 69120 Heidelberg, Germany \\
  $^5$Universit\"{a}t Heidelberg, Interdisziplin\"{a}res Zentrum f\"{u}r Wissenschaftliches Rechnen, Im Neuenheimer Feld 205, 69120 Heidelberg, Germany }
}

\date{Submitted 2021 January 18}

\pagerange{\pageref{firstpage}--\pageref{lastpage}} \pubyear{2021}

\begin{document}
\label{firstpage}
\maketitle

\begin{abstract}
We obtain two equations (following from two different approaches) for the density profile in a self-gravitating polytropic cylindrically symmetric and rotating turbulent gas disk. The adopted physical picture is appropriate to describe the conditions near to the cloud core where the equation of state of the gas changes from isothermal (in the outer cloud layers) to one of `hard polytrope', and the symmetry changes from spherical to cylindrical. On the assumption of steady state, as the accreting matter passes through all spatial scales, we show that the total energy per unit mass is an invariant with respect to the fluid flow. The obtained equation describes the balance of the kinetic, thermal and gravitational energy of a fluid element. We also introduce a method for approximating density profile solutions (in a power-law form), leading to the emergence of three different regimes. We apply, as well, dynamical analysis of the motion of a fluid element. Only one of the regimes is in accordance with the two approaches (energy and force balance). It corresponds to a density profile of a slope $-2$, polytropic exponent $3/2$, and sub-Keplerian rotation of the disk, when the gravity is balanced by the thermal pressure. It also matches with some observations and numerical works and, in particular, leads to a second power-law tail (of a slope $\sim-1$) of the density distribution function in dense, self-gravitating cloud regions.
\end{abstract}

\section{Introduction}   \label{Sec_Introduction}
Star formation in galaxies goes in molecular clouds (MCs): very cold ($T \sim 10-30$~K) objects of irregular shape consisting of molecular gas well mixed with small amounts of dust \citep[see][for a review]{BP_ea_2020}. Their internal structure is fractal within a large range of spatial scales 0.001~pc$\lesssim L \lesssim 100$~pc \citep{Elmegreen_1997, HF_2012}, with mean number densities varying from about $10^2$ cm$^{-3}$ at $L\sim$100~pc up to $>10^5$~cm$^{-3}$ in the so-called prestellar cores ($L \leq$ 0.1 pc). MC physics is complex due to the interplay of gravity, supersonic turbulence and magnetic fields; at late stages of cloud's evolution, accretion from the surrounding medium and feedback from newborn stars and supernovae plays also an important role \citep{MacLow_Klessen_2004, McKee_Ostriker_2007, KG_2016}. Gas thermodynamics in MC is generally characterized by an isothermal equation of state (EOS) $P_{\rm gas}\propto\rho^{\Gamma}$ where $\rho$ is the density and $\Gamma\approx1$. However, in denser substructures like protostellar cores the effective EOS is rather one of a `hard polytrope' with $\Gamma>1$ \citep{Fed_Ban_2015,KNW_2011}. The observable features at such small scales may also bear imprints of strong magnetic fields and/or disk rotation.

\begin{figure} 
\begin{center}
\includegraphics[width=84mm]{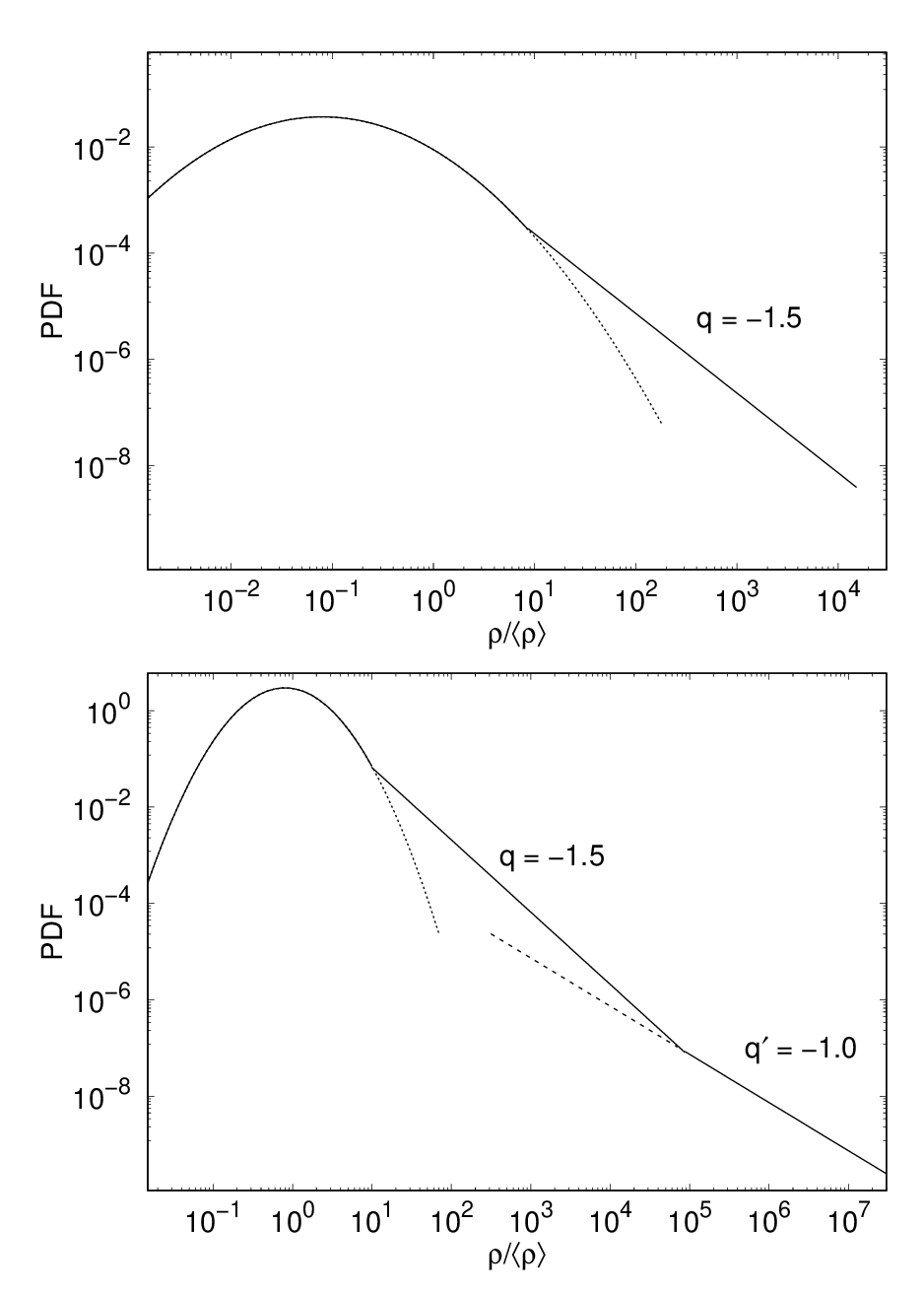}
\vspace{0.1cm}  
\caption{ The figure presents examples of analytically obtained evolved density PDFs with one (top) and two (bottom) PLTs. Please, see the text for details about the slope values. The main part of the PDF is fitted with log-normal function (dotted).}
\label{fig_PDF_PLTs}
\end{center}
\end{figure}

An increasingly popular approach to study MC physics is the analysis of the probability distribution function (PDF) of density $\rho$-PDF (which could be obtained from numerical simulations) and of column density $N$-PDF (derived from observational data and simulations). A correspondence between these statistical characteristics can be obtained assuming that the cloud is roughly spherically symmetric, with a radial density profile  $\rho(l)\propto l^{-p}$ where $l$ is a given radius. The shape of the $\rho$-PDF/$N$-PDF reflects the general structure and the evolutionary stage of the cloud. Isothermal supersonic turbulence usually dominates MC physics at scales above 1~pc -- in that case, a nearly lognormal\footnote{A Gaussian function of logdensity.} $\rho$-PDF (hereafter, simply PDF) is predicted from theoretical considerations \citep{VS_1994, Passot_VS_1998} which is also confirmed from numerous simulations \citep[e.g.,][]{Klessen_2000, Li_Klessen_MacLow_02, Kritsuk_ea_2007,Fed_ea_2010}. Main parts of $N$-PDFs derived from observational \citep{Kainu_ea_2009, Kainu_ea_2013, Lombardi_ea_2014, Schneider_ea_2015a, Schneider_ea_2015b, Schneider_ea_2016,Schneider_ea_2022} and numerical data \citep{Passot_VS_1998, Fed_ea_2008, Fed_ea_2010, Konstandin_ea_2012, Girichidis_ea_2014} have lognormal shapes as well. At smaller spatial scales (and higher densities) but well above the scales of protostellar cores, the high-density part of the PDF gradually changes its shape in the course of MC evolution from a lognormal wing to a power-law tail (PLT) $P(\rho)\propto \rho^{q}$. The typical slopes  of such PLTs: $3 \ge q \geqslant -1.5$ (Fig. \ref{fig_PDF_PLTs}, top) correspond, under the assumption of approximate spherical symmetry, to exponents of the radial density profile: $1 \le p=-3/q \le 2$ \citep[see][and the references therein]{DVK_2017}. 

Most authors attribute the emergence of a PLT to the increasing role of self-gravity in the energy balance at the corresponding small spatial scales \citep{Klessen_2000, Dib_Burkert_2005, Slyz_ea_2005, VS_ea_2008, KNW_2011, Collins_ea_2012}. Examples of the PLT evolution from numerical simulations at scales of giant MCs and of collapsing clumps are discussed in \citet{Veltchev_ea_2019}. In general, the PLT slope $q$ becomes shallower in evolving self-gravitating clouds due to formation of denser substructures (clumps, cores), with an upper limit $q\simeq-1.5$. The latter can be explained within various models: collapse of the so called singular isothermal spheres \citep{Penston_1969a, Penston_1969b, Larson_1969, Shu_1977, Hunter_1977, Whitworth_Summers_1985}, pressure-less gravitational collapse in strongly self-gravitating systems \citep{Girichidis_ea_2014},  scale-free gravitational collapse \citep{Li_2018}, collapse and dynamics of isolated gravo-turbulent cloud \citep{Jaupart_Chabrier_2020} or dynamical equilibrium between gravity and accretion \citep[][hereafter, Paper I and Paper II]{DS_2018, DS_2019}. Some numerical \citep{KNW_2011} and observational studies \citep{Schneider_ea_2015c,Schneider_ea_2022} indicate existence of a second PLT at the high-density end of the PDF associated with scales at which local collapses take place (Fig. \ref{fig_PDF_PLTs}, bottom). This PLT is usually shallower\footnote{See, however, \citet{Schneider_ea_2022} for counter-examples.}, with slope $q^\prime\simeq-1$ (corresponding to a density-profile exponent $p=3$, in the case of spherical symmetry) and hints at a slower gas accretion on the small-scale substructures in the cloud. This phenomenon still lacks a clear physical explanation; possible reasons might be rotation of contracting core(s), strong magnetic fields or change in the thermodynamic EOS \citep{KNW_2011, Schneider_ea_2015c, Donkov_ea_2021}.

\cite{Donkov_ea_2021} proposed a model in which a flatter second PLT is associated with the densest cloud parts (e.g., protostellar cores). The authors obtain an equation for the conservation of energy of a fluid element per unit mass. It is the analogue of Bernoulli equation in this context. In this study spherical symmetry (without rotation), steady state of both macroscopic and microscopic motions, and a polytropic EOS with $\Gamma>1$, are assumed. The leading order solutions correspond to four cases which have different density profiles, polytropic exponents, and energy balance equations for the fluid element. The solution with $p=3$, $(q=-1)$ and $\Gamma=4/3$ can provide an explanation of a flatter second PLT.

In this paper we continue the efforts to build an hydrodynamical model, which explain the second PL-tail in the vicinity of emerging protostar/protocluster. We consider a rotating disk (instead of static sphere) around central core (protostar), accreting material from its surrounding medium, and the fluid in the disk is in steady state in regard to macroscopic and microscopic motions, obeying also a polytropic EOS with exponent $\Gamma>1$. The analogue of Bernoulli equation for the current model is derived from Euler equation with the application of the assumptions mentioned above. It describes, as in the previous study, the conservation of energy of a fluid element per unit mass. We find an approximate solution keeping only the leading order terms. It describes three regimes characterized by different constraints for the density profiles, the polytropic exponents, the angular velocity scaling exponents, and energy balance equations for the fluid element. Performing, in addition, dynamical analysis through the force balance equation, we conclude that only one regime is in agreement with two approaches (energy and force balance equations). Namely, the second regime, where the gas pressure force balances the gravity and the rotation of the disk is sub-Keplerian. In particular it gives, as a possible solution, density profile of a slope $p=2$, which in the regarded cylindrical symmetry of the disk raises a PL-tail of a slope $q=-1$, and polytropic exponent $\Gamma=3/2$. We consider this as a possible explanation of the second PLT.

The paper has the following structure. The model of the gas medium in the vicinity of the core of the cloud is presented in Section \ref{Sec-Set up of the model}. The equation for the density profile is derived in Section \ref{Sec-Equ_rho(l)}. In more details: Section \ref{Subsec-equ of the medium} contains comments on the equations of the medium; the derivation of the general form of the equation of energy conservation for a fluid element (per unit mass) is done in Section \ref{Subsec-equ_rho(l)_general}; each term of the latter equation is given in explicit form in Section \ref{Subsec-equ_rho(l)_terms}; the equation for the density profile near to the cloud core is finally obtained in \ref{Subsec-equ_rho(l)_diff equ}. The solution is presented in Section \ref{Sec-analysis and sol}, as the energy balance equation is analysed and its solutions in three different physical regimes are obtained in Section \ref{Subsec-sol of the energy equ}, while the dynamical analysis is presented in Section \ref{Subsec-Dynamical equ}. A discussion on the obtained solutions and on the model as a whole is given in Section \ref{Sec-discussion}. Our conclusions are presented in Section \ref{Sec-conclusions}.

\section{Setup of the model}   
\label{Sec-Set up of the model}

We consider a system of molecular hydrogen and dust, which consists of a rotating spherical gaseous core and an accretion gaseous disk rotating around the core. Feedback from the new-born stars in the core are neglected. We consider the system at the earliest period of its life. The outer size of the disk we label with $l_{\rm c}$ and assume that $l_{\rm c} \gg h$ and also $l_{\rm c} \gg l_0$, where $h$ is the thickness of disk and $l_0$ is the inner size of disk, which is of the order of core's size. The core is assumed to be homogeneous, for simplicity. We, also, presume the disk is rotating differentially, and the gas in it is turbulent, self-gravitating, and accreting onto the core. We presume that there is a weak friction between the neighbouring slices (rings) of the disk, which causes the redistribution of angular momentum of the fluid elements while they are falling onto the core. Although this friction plays a role for the fluid dynamics in the disk we assume that in terms of the energy balance of each fluid element its contribution is negligible. So one can expect that the rotational kinetic energy is much larger than the kinetic energy due to the radial accretion. This is also confirmed from the results (see Section \ref{Sec-analysis and sol}) and so the presented here model is self-consistent. We also suggest there exists a radial density profile $\rho(l)$ for the disk, where $l_{\rm c} \geq l \geq l_0$ is an arbitrary radius (or scale).

We normalize the gas density and scale to $\rho_{\rm c}$ (the density at disk's edge) and $l_{\rm c}$, respectively. From now on we will work with dimensionless density $\varrho \equiv \rho/\rho_{\rm c}$ and radius $\ell \equiv l/l_{\rm c}$, where it is more appropriate. Accordingly the dimensionless density profile in the disk is $\varrho(\ell)$.

We assume the fluid in the disk to be in a steady state regarding to both: its macro-motion (the motion of fluid elements), and its micro-motion (the thermal motion of gaseous molecules), and so the temperature is locally constant in time. An arbitrary fluid element, residing at scale (radius) $\ell$, takes part in three independent motions -- accreting radially, rotating around the central core, and moving turbulently. So its velocity field, respectively, reads: $\vec{u} = \vec{u}_{\rm a} + \vec{u}_{\rm rot} + \vec{u}_{\rm t}$, where $\vec{u}_{\rm a}$ is the accretion (or radial in-fall) velocity, $\vec{u}_{\rm rot}$ is the velocity of rotation, and $\vec{u}_{\rm t}$ is the turbulent velocity, accordingly. The turbulence, we presume, to be locally homogeneous and isotropic for scales $l_{\rm c} \geq l \geq l_0$. Combining the latter presumptions with the steady state assumption, one can conclude that averaging the model (and hence the velocity field) in regard to time can be replaced by ensemble averaging in regard to an ensemble of copies of the system raised by possible different turbulent velocities of fluid elements at a fixed moment. So, if one averages turbulent velocity of a given fluid element, one obtains: $\langle \vec{u}_{\rm t} \rangle = 0$ (due to the local isotropy), and hence the averaged total velocity field reads: $\langle \vec{u} \rangle = \langle \vec{u}_{\rm a} \rangle + \langle \vec{u}_{\rm rot} \rangle = \vec{u}_{\rm a} + \vec{u}_{\rm rot}$. We note also that averaged scalar products between every two of the three velocity fields are zero:

$$ \langle \vec{u}_{\rm a} \cdot \vec{u}_{\rm t} \rangle = \langle \vec{u}_{\rm a} \rangle \cdot \langle \vec{u}_{\rm t} \rangle = 0~,$$

$$ \langle \vec{u}_{\rm rot} \cdot \vec{u}_{\rm t} \rangle = \langle \vec{u}_{\rm rot} \rangle \cdot \langle \vec{u}_{\rm t} \rangle = 0~,$$

and

$$ \langle \vec{u}_{\rm a} \cdot \vec{u}_{\rm rot} \rangle = \vec{u}_{\rm a} \cdot \vec{u}_{\rm rot} = 0~,$$
hence the three averaged velocity fields are orthogonal to each other. From now on we will use, where it is appropriate, dimensionless velocity fields substituting $u$ for $v\equiv u/c_{\rm s}$, where $c_{\rm s}$ is the isothermal sound speed explained below.

The entire system (core plus disk) is a central part of a larger cloud which material obeys radial symmetry and has its own density profile (probably different from the profile in disk). This cloud consists of molecular hydrogen, and is isothermal, self-gravitating and turbulent. It accretes matter from its surroundings and this material moves through the cloud shells to reach the central part where our system is located. We presume that the cloud shells are in the steady state, as well, regarding accretion, turbulence, and thermodynamics. These shells cause a constant gravitational potential $\varPhi^{\rm ext}$ in the volume where our system resides, and therefore it does not affect the motion of fluid elements in the disk (for further details see Paper II).

Regarding thermodynamics of the gas in the disk we assume a polytropic equation of state (EOS):

\begin{equation}
	\label{equ_polytropic EOS}
	P_{\rm gas}=P_0 (\rho/\rho_{\rm c})^\Gamma~,~~~\Gamma>1~~.
\end{equation}
It reflects the typical assumption for the medium in the densest part of the cloud, near to the core \citep{Fed_Ban_2015}. When $\Gamma=1$ this EOS reduces to the isothermal one $P_{\rm gas}=c_{\rm s}^2 \rho$, with $c_{\rm s}$ being the speed of sound \citep{Fed_Ban_2015} at some fixed temperature ($10 {\rm K} \leq T \leq 30 {\rm K}$), which characterizes the outer cloud matter. From the comparison of the two equations we see that $P_0=c_{\rm s}^2 \rho_{{\rm c}}$. With this choice we can connect continuously the two EOS of the molecular gas: the isothermal one (for the outer shells) and the `hard polytropic' one (for the disk).

We neglect magnetic fields and stellar feedback (in spite of they proved to be of great importance; see for instance: \cite{Lebreuilly_ea_2021,Lebreuilly_ea_2023,Hennebelle_ea_2022}), and we note that this is a considerable simplification of the reality. Also the disk is presumed to be very thin, although, from simulations (see for example \cite{Bate_2018,Lebreuilly_ea_2021}) and observations (see for example \cite{Andrews_ea_2018a,Andrews_ea_2018b,Sheehan_ea_2022a,Sheehan_ea_2022b,Tobin_ea_2020a,Tobin_ea_2020b}), there are many examples where protostellar disks look fluffy. Hence we deal with a very simple hydrodynamic model of protostellar system, but we consider this is admissible in regard to our efforts to obtain the density profile $\varrho(\ell)$ (and hence the PDF) near the core, in first approximation.

\section{Derivation of the equation for density profile near to the core}
\label{Sec-Equ_rho(l)}

\subsection{The fluid equations}
\label{Subsec-equ of the medium}
We start with the equations of the medium, as a strong base in our aim to obtain the density profile $\varrho(\ell)$ near to the core. They read as follow:

\begin{itemize}
\item[-] The Euler equations

\begin{equation}
\label{equ_cont}
\frac{\partial\rho}{\partial t} + \nabla\cdotp(\rho\vec{u})=0~~,
\end{equation}

\begin{equation}
\label{equ_N-St}
\frac{\partial\vec{u}}{\partial t} + \vec{u}\cdotp\nabla\vec{u} = -\frac{1}{\rho }\nabla P_{\rm gas} - \nabla\varPhi ~~,
\end{equation} 
where the first one is the mass conservation law, while the second one is the equation of motion for a fluid element (here $\varPhi$ is the total gravitational potential). The terms of equation (\ref{equ_N-St}) will be specified in Section \ref{Subsec-equ_rho(l)_terms}. The dissipative terms and the friction between neighbouring rings have been neglected. The latter was also mentioned in Section \ref{Sec-Set up of the model}. The former is justified by our assumption of statistical equilibrium in regard to turbulence.

\item[-] The polytropic EOS (\ref{equ_polytropic EOS}) of the gas introduced in the previous Section \ref{Sec-Set up of the model}.

\item[-] The equation which determines the gravitational potential (the Poisson equation), according to the given density distribution

\begin{equation}
\label{equ_Pois}
\Delta\varPhi=4\pi G \rho~~.
\end{equation}
  
\end{itemize}

\subsection{General form of the equation of energy conservation for a moving fluid element}
\label{Subsec-equ_rho(l)_general}

The equation for $\varrho(\ell)$ must take into account the assumptions for the symmetry and the physics of the idealized object that we describe.

First we remind the reader, according to \cite{Donkov_ea_2021}, the derivation of the equation in its general form (here equation \ref{dE=0}). The derivation is independent of the symmetry of the system in regard. This repetition we considered as very useful for readability of the text. It is as follows:

\begin{itemize}
\item[i)] Starting with some algebra of polytropic EOS: From $\nabla P_{\rm gas}/\rho=c_{\rm s}^2 (\rho_{\rm c}/\rho) \nabla (\rho/\rho_{\rm c})^\Gamma=c_{\rm s}^2 \varrho^{-1} \nabla \varrho^\Gamma=...=c_{\rm s}^2 (\Gamma/(\Gamma-1)) \nabla \varrho^{\Gamma-1}$ we obtain	
\[ \frac{\partial\vec{u}}{\partial t} + \vec{u}\cdotp\nabla\vec{u} = -c_{\rm s}^2\frac{\Gamma}{\Gamma-1}\nabla\varrho^{\Gamma-1} - \nabla\varPhi ~~.\]
	
\item[ii)] Then we multiply the above equation by an infinitesimal displacement $d\vec{r}=\vec{u}dt$ directed to the vector field $\vec{u}$:
\[ \vec{u}\cdotp\frac{\partial\vec{u}}{\partial t}dt + (\vec{u}\cdotp\nabla\vec{u})\cdotp\vec{u}dt = \]
\[ -c_{\rm s}^2 \frac{\Gamma}{\Gamma-1}(\nabla\varrho^{\Gamma-1})\cdotp d\vec{r} - (\nabla\varPhi)\cdotp d\vec{r} ~~,\]
and obtain for the left hand side the following expression 
\[  \vec{u}\cdotp\frac{\partial\vec{u}}{\partial t}dt + (\vec{u}\cdotp\nabla\vec{u})\cdotp\vec{u}dt = \frac{d}{dt}(u^2/2)dt = d(u^2/2)~~.\]
	
\item[iii)] At that point we introduce dimensionless variables $v^2\equiv u^2/c_{\rm s}^2$  and $\phi\equiv\varPhi/c_{\rm s}^2$ and hence arrive at:
\[ d(v^2/2) = \]
\[ -\frac{\Gamma}{\Gamma-1}(\nabla\varrho^{\Gamma-1})\cdotp d\vec{r} - (\nabla \phi)\cdotp d\vec{r} ~~.\]
	
\item[iv)] With the substitutions $d(\varrho^{\Gamma-1}) = (\partial (\varrho^{\Gamma-1})/\partial t)dt + (\nabla\varrho^{\Gamma-1})\cdotp d\vec{r}$ and $d\phi = (\partial\phi/\partial t)dt + (\nabla\phi)\cdotp d\vec{r}$, we obtain finally:

\[ d(v^2/2) = \] 
\[ =-\frac{\Gamma}{\Gamma-1}[d(\varrho^{\Gamma-1}) - (\partial (\varrho^{\Gamma-1})/\partial t)dt] - [d\phi - (\partial\phi/\partial t)dt] ~~.\]
\end{itemize}

Up to this point the considerations were rather general. From now on, we will focus on our idealized object and on the motion of the fluid elements through the disk, in particular. We use brackets to designate the quantities which belong to it and are received after ensemble averaging with respect to the chaotic turbulent motion of the fluid element in a given space volume. By their use, the equation obtained at step iv) above can be rewritten:
\begin{eqnarray}
\begin{aligned}
d\langle v^2/2 \rangle = \nonumber\\
-\frac{\Gamma}{\Gamma-1}[d\langle \varrho^{\Gamma-1} \rangle - (\partial \langle \varrho^{\Gamma-1} \rangle/\partial t)dt] - [d\langle \phi \rangle - (\partial \langle \phi \rangle/\partial t)dt]~.
\end{aligned}
\end{eqnarray}

Here we apply the assumption for steady state, from which stems that $\partial \langle \varrho^{\Gamma-1} \rangle/\partial t = 0$, and $\partial \langle \phi \rangle/\partial t = 0$. This leads to:
\begin{equation}
\label{dE=0}
d\bigg[\langle v^2/2 \rangle + \frac{\Gamma}{\Gamma-1}\langle \varrho^{\Gamma-1} \rangle + \langle \phi \rangle \bigg] =0~.
\end{equation}

We assumed that turbulence is locally homogeneous and isotropic. Hence, it does not affect the ensemble averaged motion of the fluid elements. As a result, it is a superposition of radial in-fall and rotation about the centre. Turbulence, however, must be taken into account when the kinetic energy term $v^2/2$ is calculated, because the velocity $\vec{v}$ is first squared and then averaged. So one has: $v^2 = v_{\rm a}^2 + v_{\rm rot}^2 + v_{\rm t}^2$. Moreover, the system in regard is described by an averaged density profile, which is closely connected with the PDF, which, in its turn, is averaged by assumption (see Paper I). Then, $\langle\varrho^{\Gamma-1}\rangle=\varrho^{\Gamma-1}$. Following the above reasoning one gets an equivalent form of equation (\ref{dE=0}):

\begin{equation}
\label{dE=0_final}
d\bigg[\langle v_{\rm a}^2/2 + v_{\rm rot}^2/2 + v_{\rm t}^2/2 \rangle + \frac{\Gamma}{\Gamma-1}\varrho^{\Gamma-1} + \langle \phi \rangle \bigg]=0~.
\end{equation}

\subsection{Explicit form of the terms in equation (\ref{dE=0_final})}
\label{Subsec-equ_rho(l)_terms}

The derivation of the explicit form of the terms in equation (\ref{dE=0_final}) is done as follows. 

\subsubsection{Kinetic term}
\label{Subsubsec-kinetic term}

As shown in the previous Section \ref{Subsec-equ_rho(l)_general}, the kinetic energy term reads:
\begin{equation}
\label{v=va+vrot+vt}
\langle v^2 \rangle = \langle v_{\rm a}^2 \rangle + \langle v_{\rm rot}^2 \rangle + \langle v_{\rm t}^2 \rangle~,
\end{equation}
where $\langle v_{\rm a}^2 \rangle$ is the accretion kinetic energy per unit mass, $\langle v_{\rm rot}^2 \rangle$ is the rotational kinetic energy per unit mass, and $\langle v_{\rm t}^2 \rangle$ is the turbulent kinetic energy per unit mass. 

In general, turbulent velocity fluctuations in molecular clouds obey a power-law scaling relation $u=u_0 (L/1~\mathrm{pc})^{\beta}$ \citep{Larson_1981, Padoan_ea_2006, Kritsuk_ea_2007, Fed_ea_2010}, with some normalizing factor $u_0$ and scaling index $0\leq\beta\leq1$. We adopt this relation and assume $\langle v_{\rm t}^2 \rangle$ depends on the scale $\ell$ as:
\begin{eqnarray}
\label{vt2-p(s)}
\langle v_{\rm t}^2 \rangle = \frac{u_0^2}{c_{\rm s}^2}\bigg(\frac{l}{pc}\bigg)^{2\beta} \equiv T_0 \ell^{2\beta} ~,
\end{eqnarray}
where the constant $T_0 = (u_0^2/c_{\rm s}^2)(l_{\rm c}/pc)^{2\beta}$ is the ratio of the turbulent kinetic energy per unit mass of a fluid element to the thermal energy per unit mass, at the largest scale of the disk; and $0 \leq \beta \leq 1$ is a scaling exponent of the turbulent flow.

The accretion kinetic term $\langle v_{\rm a}^2 \rangle$ in its explicit form can be got from the continuity equation, as it is shown in Paper I (see Section 3.3 there). Here we apply a similar treatment but adopting cylindrical symmetry, which is more appropriate for the disk. The main steps, in cylindrical coordinates $(l, \varphi, z)$, are as follows. The continuity equation (Eq. \ref{equ_cont}) is averaged with respect to the ensemble of the micro-states of the system, caused by the turbulence:
\[ \frac{\partial\rho}{\partial t} + \nabla\cdotp\langle\rho\vec{u}\rangle=0~.\]
Applying the steady state assumption one obtains $\partial\rho/\partial t =0$, and hence:
\[ \nabla\cdotp\langle\rho\vec{u}\rangle = \nabla\cdotp\rho\langle\vec{u}\rangle = 0 ~.\]
If the motion of the fluid element is averaged the turbulent velocity vanishes (i.e: $\langle \vec{u} \rangle = \langle \vec{u}_{\rm a} \rangle + \langle \vec{u}_{\rm rot} \rangle = \vec{u}_{\rm a} + \vec{u}_{\rm rot} = (u_{\rm a}, u_{\rm rot}, 0)$). And also taking into account that the density profile is averaged by assumption, we further obtain:
\[ \frac{1}{l} \frac{\partial}{\partial l} (l\rho u_{\rm a}) + \frac{1}{l} \frac{\partial}{\partial \varphi} (\rho u_{\rm rot}) + \frac{\partial}{\partial z} (\rho u_{\rm z}) = \frac{1}{l} \frac{\partial}{\partial l} (l\rho u_{\rm a}) = 0 ~, \]
where we implicitly assume that $\rho$, $u_{\rm a}$ and $u_{\rm rot}$ depend only on the scale $l$, but not on the polar angle $\varphi$ and on the $z-$coordinate. Finally, one obtains the equation:
\[ l\rho u_{\rm a} = {\rm const} ~,\]
and, hence, a formula for $\langle v_{\rm a}^2 \rangle$:
\begin{equation}
\label{va2-rho(l)}
\langle v_{\rm a}^2\rangle = A_0 \varrho(\ell)^{-2} \ell^{-2}~.
\end{equation}
Here the dimensionless coefficient $A_0$ is obtained as the ratio of the accretion kinetic energy term at the disk edge to the thermal kinetic energy, at isothermal EOS, per unit mass.

The rotation kinetic energy $\langle v_{\rm rot}^2 \rangle$ can be written in the form: $\langle v_{\rm rot}^2 \rangle = A_{\rm rot} \omega(\ell)^2 \ell^2$, where $\omega(\ell)$ is the dimensionless angle velocity at scale $\ell$, and $A_{\rm rot}$ is the ratio of the rotational kinetic energy at the disk's edge to the thermal energy (at isothermal EOS), per unit mass. Here we presume an explicit formula for $\omega(\ell)$ in the form of power-law: $\omega(\ell) = \ell^{-c}$, where $c>0$ is a positive exponent. The latter means that the closer is a ring with radius $\ell$ to the centre, the larger is its angular velocity $\omega(\ell)$ (remember that $\ell<1$). Hence the rotational kinetic energy reads:

\begin{equation}
	\label{vrot2-l}
	\langle v_{\rm rot}^2 \rangle = A_{\rm rot} \ell^{2(1-c)} ~.
\end{equation}

Therefore, finally, the full kinetic energy term is as follows:

\begin{equation}
	\label{v2-l}
	\langle v^2 \rangle = A_0 \varrho(\ell)^{-2} \ell^{-2} + T_0 \ell^{2\beta} + A_{\rm rot} \ell^{2(1-c)} ~.
\end{equation}

\subsubsection{Gravity potential term}
\label{Subsubsec-gravity potential term}

In order to calculate the gravitational term $\langle \phi \rangle$ in equation (\ref{dE=0_final}) we consider the disk as built of narrow rings with masses $\Delta M$ and radii $l$. Then an arbitrary ring causes a potential $\phi(l,a)$ at a distance $a$ from the centre. There are two possible approximations\footnote{This is only valid in the disk mid-plane, hence the thin disk approximation.}:
\begin{itemize}
	\item [i)] If $l>a$, then:
	\[ \phi(l,a) = - G \frac{\Delta M}{l} \bigg[ 1 + \frac{1}{4}\bigg( \frac{a}{l} \bigg)^2 +\frac{9}{64}\bigg( \frac{a}{l} \bigg)^4 + ...  \bigg] \approx \]
	\[ \approx - G \frac{\Delta M}{l} ~; \]
	
	\item [ii)] If $l<a$, then:
	\[ \phi(l,a) = - G \frac{\Delta M}{a} \bigg[ 1 + \frac{1}{4}\bigg( \frac{l}{a} \bigg)^2 +\frac{9}{64}\bigg( \frac{l}{a} \bigg)^4 + ...  \bigg] \approx \]
	\[ \approx - G \frac{\Delta M}{a} ~. \] 
\end{itemize}
In both cases we confine the considerations only to the leading order term in regard to simplicity, and due to the series converge fast. Moreover, the higher order terms will raise subdominant terms in the explicit form of equation (\ref{dE=0_final}). Hence, if the fluid element resides at scale $a$, the first case will help us to calculate the gravitational potential caused by the rings located in disk plane at larger radii, and the second case to account for those located below it.

Making use of the above treatment we are able to write the explicit formula for the gravitational term:
\begin{equation}
	\label{phi-general}
	\langle \phi \rangle = -G_0 \int_{\ell}^{1} \varrho(\ell^{'}) d\ell^{'} - G_0 \frac{1}{\ell} \int_{\ell_0}^{\ell} \varrho(\ell^{'}) \ell^{'} d\ell^{'} - G_1 \frac{1}{\ell} ~,
\end{equation}
where the first term is caused by the rings above the fluid element, the second term is due to the rings below the fluid element, and the last term presents the gravity of the spherical core. The dimensionless coefficients $G_0$ and $G_1$ are defined as follows: $G_0 \equiv 2\pi G l_{\rm c} h \rho_{\rm c} / c_{\rm s}^2$ and it means (with a precision to a factor of order unity) the ratio of the gravitational energy of homogeneous disk (in the first order approximation) to the thermal energy, for an isothermal EOS, per unit mass; $G_1 \equiv G M_0 / l_{\rm c} c_{\rm s}^2$ and this coefficient has the same physical meaning but for the central spherical homogeneous core.

To calculate the gravitational term in an explicit form one needs of formula for $\varrho(\ell)$. In Section \ref{Sec-Set up of the model} we assume that there exists a density profile for the disk, and now we extend it to a power-law form: $\varrho(\ell) = \ell^{-p}$, where $p>0$ is a positive exponent \footnote{This immediately leads to an explicit formula for the accretion kinetic term: $\langle v_{\rm a}^2\rangle = A_0 \ell^{2(p-1)}$}. The latter raises three possible cases:

\begin{itemize}
	\item [1)] If $p=1$, then one has:
	\[ \langle \phi \rangle = G_0 \ln(\ell) - G_0 [1-(\ell_0/\ell)] - G_1 \ell^{-1} ~; \]
	
	\item [2)] If $p=2$, then one has:
	\[ \langle \phi \rangle = G_0 (1-\ell^{-1}) - G_0 \ell^{-1} \ln(\ell/\ell_0) - G_1 \ell^{-1} ~; \]
	
	\item [3)] If $p\neq1$ or $2$, then one has:
	\[ \langle \phi \rangle = - \frac{G_0}{1-p} (1-\ell^{1-p}) - \frac{G_0}{2-p} \ell^{1-p} [1-(\ell_0/\ell)^{2-p}] - G_1 \ell^{-1} ~. \]
\end{itemize}
Although there exist three possible cases for $p>0$, the most probable range for the density exponent is: $2\geq p \geq1$ (see simulation by \cite{Bate_2018} and \cite{KNW_2011}; and observations by \cite{Schneider_ea_2015c,Schneider_ea_2022}). And if we regard the scales near to the core: $\ell\gtrsim\ell_0$, then in all the three cases for $\langle \phi \rangle$ the leading order term(s) scales with $\ell^{-1}$. One can see this taking into account that $0<\ell<1$ and the leading order term(s) must have the smallest exponent. So making use of the latter considerations, in these three possible cases we have:

\begin{itemize}
	\item [1)] If $p=1$, then:
	\[ \langle \phi \rangle \approx - G_1 \ell^{-1} ~, \]
	the first term is subdominant and the second vanishes;
	
	\item [2)] If $p=2$, then:
	\[ \langle \phi \rangle \approx - G_0 \ell^{-1} - G_1 \ell^{-1} ~, \]
	the second term vanishes;
	
	\item [3)] If $p\neq1$ or $2$, then:
	\[ \langle \phi \rangle \approx - G_1 \ell^{-1} ~, \]
	the first term is subdominant and the second is too (because the second addend in the parentheses $\ell_0^{2-p}\ell^{-1} \sim \ell^{1-p} < \ell^{-1}$ for $\ell\gtrsim\ell_0$ and $2>p>1$ ).
\end{itemize}
To simplify further considerations we can present all the three cases with the term: $-G^{\rm (i)}\ell^{-1}$, where $G^{\rm (1)} = G_1$ in the first case; $G^{\rm (2)} = G_0 + G_1$ in the second case; and $G^{\rm (3)} = G_1$ in the third case, accordingly.

\subsection{Derivation of the equation for $\varrho(\ell)$}
\label{Subsec-equ_rho(l)_diff equ}

At last Eq. (\ref{dE=0_final}) can be written in a appropriate form, allowing us to derive an equation for the density profile:
\begin{eqnarray} \label{dE/dl=0_rho}
\begin{aligned}	
\frac{d}{d\ell}\Bigg[ A_0 \ell^{2(p-1)} + T_0 \ell^{2\beta} + A_{\rm rot} \ell^{2(1-c)} \\
+2\frac{\Gamma}{\Gamma-1}\ell^{-p(\Gamma-1)} - 2G_0 \int_{\ell}^{1} (\ell^{'})^{-p} d\ell^{'} \\
- 2G_0 \frac{1}{\ell} \int_{\ell_0}^{\ell} (\ell^{'})^{1-p} d\ell^{'} - 2G_1 \frac{1}{\ell} \Bigg]=0~,
\end{aligned}
\end{eqnarray}
where we have substituted the total differential `$d$' for the derivative `$d/d\ell$', because all the terms depend only on $\ell$. The expression in the brackets in Eq. (\ref{dE/dl=0_rho}) is the total energy per unit mass of a fluid element. Labelling it by $E_0$, one receives:
\begin{eqnarray} \label{E_rho}
\begin{aligned}
A_0 \ell^{2(p-1)} + T_0 \ell^{2\beta} + A_{\rm rot} \ell^{2(1-c)} \\ +2\frac{\Gamma}{\Gamma-1}\ell^{-p(\Gamma-1)} - 2G_0 \int_{\ell}^{1} (\ell^{'})^{-p} d\ell^{'} \\ 
- 2G_0 \frac{1}{\ell} \int_{\ell_0}^{\ell} (\ell^{'})^{1-p} d\ell^{'} - 2G_1 \frac{1}{\ell} = E_0 = \mathrm{const}~.
\end{aligned}
\end{eqnarray}
This is the equation for the dimensionless density profile $\varrho(\ell) = \ell^{-p}$. As one can easily see this equation depends, also, on two more unknown parameters (except for $p$): the hard polytropic exponent $\Gamma>1$, and exponent $c>0$, determining the power-law of differential rotation. The analysis and possible solutions of Eq. (\ref{E_rho}) we present in the next Section \ref{Sec-analysis and sol}.

To analyse the equation (\ref{E_rho}) easier let us replace all the three gravitational terms with the term: $-2G^{\rm (i)}\ell^{-1}$, which implies that we confine our considerations to the range: $2\geq p \geq1$ for the density profile exponent, and to the scales near the core: $\ell\gtrsim\ell_0$. The obtained equation reads:
\begin{eqnarray} \label{Ealg_rho}
	\begin{aligned}
		A_0 \ell^{2(p-1)} + T_0 \ell^{2\beta} + A_{\rm rot} \ell^{2(1-c)} \\
		+ 2\frac{\Gamma}{\Gamma-1}\ell^{-p(\Gamma-1)} - 2G^{\rm (i)}\ell^{-1} = E_0=\mathrm{const}~.
	\end{aligned}
\end{eqnarray}

\section{Study of the derived equation for density profile}
\label{Sec-analysis and sol}

\subsection{Study of the equation (\ref{Ealg_rho})}
\label{Subsec-sol of the energy equ}

A general approach consists in obtaining an approximate solution of the equation (\ref{Ealg_rho}), where only the leading order terms does matter. Remember that term(s) with smallest exponent is/are dominant, because $\ell$: $0<\ell_0 \lesssim \ell <1$. When one rejects the subdominant term(s), the rest must consist at least of two leading terms with opposite signs, in order to balance the equation. If the dominant term is only one, or the dominant terms have one and the same sign, then the equation does not have a solution.

Let us assume that the gravitational term cannot be neglected, i.e. it is of leading order\footnote{Strictly speaking the solutions that we obtain are valid only close to the central core, since we require that $\ell\gtrsim \ell_0$. If this condition is not satisfied then the role of the subdominant terms has to be taken into account.}. Then the dominant power of $\ell$ is ``$-1$''. Obviously the turbulent kinetic term $T_0 \ell^{2\beta}$ is not important for the solution. Since $2\geq p \geq1$, the accretion kinetic term can not be important, also. The gravitational pull must be balanced by gas thermal pressure and/or centrifugal force. Here one can see three possible regimes, for which the equation (\ref{Ealg_rho}) has an approximate solution:
\begin{itemize}
	\item [I)] Only the centrifugal force balances the gravity.
	Then $2(1-c)=-1$ and $-p(\Gamma-1)>-1$. Hence $c=3/2$ and $1<\Gamma<1+1/p$. So, if $2\geq p \geq1$, then $1<\Gamma<2$. The corresponding energy balance equation is:
	\[ A_{\rm rot} - 2G^{\rm (i)} \approx 0 ~.\]
	
	\item [II)] Only the thermal pressure balances the gravity.
	Then $2(1-c)>-1$ and $-p(\Gamma-1)=-1$. Therefore $c<3/2$ and $\Gamma=1+1/p$. Since $2\geq p \geq1$, then $3/2\leq\Gamma\leq2$. The corresponding energy balance equation is:
	\[ 2\Gamma/(\Gamma-1) - 2G^{\rm (i)} \approx 0 ~.\]
	
	\item [III)] Both terms balance the gravity.
	Then $2(1-c)=-1$ and $-p(\Gamma-1)=-1$. Therefore $c=3/2$ and $3/2\leq\Gamma\leq2$. The corresponding energy balance equation is:
	\[ A_{\rm rot} + 2\Gamma/(\Gamma-1) - 2G^{\rm (i)} \approx 0 ~.\]
\end{itemize}
It is worth to note that the possibilities: $2(1-c)<-1$ and/or $-p(\Gamma-1)<-1$ are forbidden, since if one of them or both are realized then the gravity term does not matter, and hence the equation does not have a solution. Therefore this analysis restricts the possible values for $c$ and $\Gamma$, as it follows: $0< c \leq3/2$ and $1< \Gamma \leq2$.

\subsection{Dynamical analysis of the obtained results}
\label{Subsec-Dynamical equ}

We continue our study with an analysis of the obtained results in the previous Section \ref{Subsec-sol of the energy equ}. Here we change the point of view and regard the equations of motion for a fluid element in polar coordinates (equation trough the $z-$axis is not regarded, because through $z$ only the turbulent motion is relevant and as it commented in Section \ref{Sec-Set up of the model} its averaged contribution is zero). We neglect the friction and dissipation forces according to the model assumptions. We account only for the gravitational and pressure-gradient forces. Then the gravity must be balanced by centrifugal and pressure forces, in radial direction. In the tangential to the circle orbit direction there are no forces. Therefore the normalized equations of motion are:

\begin{equation}
	\label{equ-radial equ of motion}
	\ddot{\ell} - A_{\rm rot}\ell\omega^2 = -G^{(i)}\frac{1}{\ell^2} + p\Gamma \ell^{-p(\Gamma-1)-1}~,
\end{equation}
and

\begin{equation}
	\label{equ-tangential equ of motion}
	\frac{d}{dt}(\ell^2 \omega) \approx 0~,
\end{equation}
where the first equation is trough the disk radius and the second is in the tangential direction, accordingly. Make use from the scaling relation for the angular velocity $\omega(\ell)=\ell^{-c}$, one obtains that the equation (\ref{equ-tangential equ of motion}) gives: $(2-c)\ell^{1-c}\dot{\ell}\approx0$, hence $\dot{\ell}\approx0$, and finally $\ddot{\ell}\approx0$. Using the latter we obtain from equation (\ref{equ-radial equ of motion}) the following force balance equality:

\begin{equation}
	\label{equ-force balance}
	A_{\rm rot} \ell^{1-2c} + \frac{1}{p\Gamma} \ell^{-p(\Gamma-1)-1} \approx G^{(i)} \ell^{-2}~.
\end{equation}

One can easily see that the equation (\ref{equ-force balance}) reproduces the approximation obtained in the previous Section \ref{Subsec-sol of the energy equ}, but with different energy balance equations:

\begin{itemize}
	\item [1)] If only the centrifugal force balances the gravity, then $1-2c=-2$, and hence $c=3/2$; moreover $-p(\Gamma-1)-1>-2$, and therefore $1<\Gamma<2$; the corresponding energy balance is:
	\[ A_{\rm rot} - G^{\rm (i)} \approx 0 ~.\]
	\item [2)] If only the gas pressure balances the gravity, then $-p(\Gamma-1)-1=-2$ and $1-2c>-2$, and hence $3/2\leq\Gamma\leq2$ and $0<c<3/2$; the corresponding energy balance is:
	\[ p\Gamma - G^{\rm (i)} \approx 0 ~.\]
	\item [3)]  If both terms on the left hand side balance the gravity, then $c=3/2$ and $3/2\leq\Gamma\leq2$; and the corresponding energy balance is:
	\[ p\Gamma + A_{\rm rot} - G^{\rm (i)} \approx 0 ~.\]
\end{itemize}

The dynamical approach in this Section must be consistent with the energy equation which we solved in the previous Section \ref{Subsec-sol of the energy equ}, therefore the regimes I and 1, II and 2, III and 3, must be in agreement, respectively. Solving the energy balance equations in the regimes I and 1 together, we obtain that the only solution is: $A_{\rm rot}=0,~G^{(i)}=0$, which is \modify{trivial and} not realistic. Solving simultaneously the energy balance equations in the regimes II and 2, we obtain $G^{(i)}=1+p$ together with $\Gamma=1+1/p$. And, finally, in the regimes III and 3, the energy balance equations give: $A_{\rm rot}=0,~G^{(i)}=1+p$, which is unrealistic like in the first regime. So one might conclude that the only reasonable approximate solution is the second regime, when the gas pressure force balances the gravity and the rotation of the disk is sub-Keplerian.

For better readability of the obtained results in this paragraph we summarized them in table \ref{table-summary of the results}.

\begin{table*}
	\begin{minipage}{140mm}
		\centering
		\caption{Summary of the results}
		\centering\small
		\begin{tabular}{cccc}
			\hline 
			regimes & First  & Second & Third  \\
			\hline
			density profile & $1\leq p \leq 2$ & $1\leq p \leq 2$ & $1\leq p \leq 2$\\
			\\
			PLT slope & $-2\leq q \leq -1$ & $-2\leq q \leq -1$ & $-2\leq q \leq -1$\\
			\\
			angular vel. scaling exp. & $3/2$  & $< 3/2^*$ & $3/2$\\
			\\
			polytropic exponent & $1<\Gamma<2$ & $3/2\leq\Gamma\leq2$ & $3/2\leq\Gamma\leq2$\\
			\\
			Energy balance (from Eq. (15)) & $A_{\rm rot}=2G^{\rm (i)}$  &$\frac{2\Gamma}{\Gamma-1}=2G^{\rm (i)}$& $A_{\rm rot}+\frac{2\Gamma}{\Gamma-1}=2G^{\rm (i)}$ \\
			\\
			Energy balance (from Eq. (18)) & $A_{\rm rot}=G^{\rm (i)}$  & $p\Gamma=G^{\rm (i)}$ &$p\Gamma+A_{\rm rot}=G^{\rm (i)}$ \\
			\\
			Agreement & $A_{\rm rot}=0$  & $G^{\rm (i)}=1+p$& $A_{\rm rot}=0$\\
			& $G^{\rm (i)}=0$ & $\Gamma=1+1/p$ & $G^{\rm (i)}=1+p$\\
			\\
			Conclusion &trivial & \it{fiducial} & non-realistic\\
			\hline
			$^*$ sub-Keplerian case
		\end{tabular} 	\label{table-summary of the results}   
	\end{minipage}
\end{table*}

\section{Discussion}
\label{Sec-discussion}

\subsection{Towards a fiducial model}
\label{Subsec-discussion on the model and results}

In the presented work we suggest a model for a thin disk around protostar (the central core). Our goal is to describe the most inner part of a protostellar cloud (the larger cloud in which is embedded the regarded system, see Section \ref{Sec-Set up of the model}) in a fiducial way, so that to obtain the PDF of density of this subsystem (namely the second PL-tail of the distribution) which is observed in simulations and observations.

Building the model we account for the self-gravity of the disk and core (due to the assumption for radial symmetry of the outstanding material its gravitational field does not affect the motion of fluid elements in the disk), for the thermodynamics, turbulence, and accretion. Our model is purely hydrodynamical (no magnetic fields are regarded), the disk is thin (its perpendicular size $h\ll l$ is much less than its radius at all regarded scales), and the feedback from new-born stars is also neglected. Due to the latter suppositions our treatment is highly idealized, comparing to the results from simulations and observations, where it is clear that magnetic fields essentially concern the physics and morphology of protostellar disks. Also the observed disks are rarely thin, usually they are fluffy. Finally the new-born stars in the centre produce, in addition, outflows and radiation, which make the physical picture more complicated. Furthermore we neglect the viscosity of the fluid and friction between the neighbouring rings of the disc in the considered equations. So the suggested model is very simple and reflects only a part of the real state in which resides the very inner part of a protostellar cloud. Nevertheless, given the highly difficult physics in the real protostellar disks, our approach looks justifiable as a first approximation.

Applying together the equation (\ref{Ealg_rho}) for conservation of the total energy of a fluid element and equation (\ref{equ-force balance}) for the force balance, we concluded in the previous Section \ref{Subsec-Dynamical equ}, that there exists only one approximate solution of these equations, which corresponds to the regime where the gravitational force is balanced by the gas pressure force and the rotation of the disk is sub-Keplerian (so the scaling index of the angular velocity is in the range: $0<c<3/2$). Accordingly, the energy balance equation for a fluid element reads: $ p\Gamma - G^{\rm (i)} \approx 0$, and the polytropic exponent is given by the formula: $\Gamma = 1+1/p$, hence $G^{(i)}=1+p$. Taking into account that we suppose the density profile exponent $p$ to be in the range $1\leq p \leq2$, this gives the ranges for $\Gamma$ and $G^{(i)}$. They are $2\geq \Gamma \geq 3/2$ and $2\leq G^{(i)} \leq3$, respectively.

We continue our analysis of the solutions comparing them with realistic power-law PDFs, obtained from numerical experiments. Our aim is to restore the density PDF near to the cloud core, where the EOS changes from isothermal to `hard polytrope' one, and protostellar disk forms. To compare the model with observations and simulations, one has to account for the 2D symmetry of the disk. Then the slope $q$ of the density PL-tail is related to the density exponent $p$, trough the formula: $q=-2/p$ (\cite{DVK_2017}, and the references therein). Then, if $1\leq p \leq2$, hence $-2\leq q \leq-1$. The slope $q=-1$ corresponds to density exponent $p=2$, and therefore $\Gamma=3/2$, $G^{(i)}=3$. Some observations \citep{Schneider_ea_2015c} and simulations \citep{KNW_2011, Veltchev_ea_2019, Marinkova_ea_2021} indicate the emergence of a second PLT of the PDF, with slope $q\sim-1$. In the hydrodynamical simulation by \cite{Bate_2018}, where more than 100 protostellar disks are studied, the author claim that the radial surface density profiles of isolated disks typically scaling as $\Sigma\sim\ell^{-1}$. This can be interpreted in our case as $\varrho=\ell^{-1}$, due to the constant vertical size $h$ of the disk. Therefore the PDF slope must be $q=-2$ (accordingly $\Gamma=2$ and $G^{(i)}=2$), which is observed in some clouds by \cite{Schneider_ea_2022}.

We, also, have to mention that \cite{Bate_2018}, in their fig.16, present a cumulative statistics of the studied disks in regard to the radial surface density profile exponent (in our case the density exponent $p$), corresponding to different ranges for the radius (in regard to the mass-fraction of the disk, containing in these ranges). The main that one can conclude, from their graphics, is that the scaling exponent is not constant for all the radii in regard. Also, for all the disks $p=2$ is an upper limit (and about 80 percent have for upper limit the value $p=1$), and the values $0\leqslant p<1$ are presented, too. The latter means that one may expect the slopes $q<-2$, for the second PL-tail of the density PDF, to be observed.

For our model, the above results, would mean that the range for $p$ can be extended to $0<p\leqslant2$. Hence in the case of the only valid approximate solution the gas polytropic exponent can take values in the interval: $3/2\leqslant\Gamma<\infty$, and the energy coefficient accounting for the self-gravity will be in the range: $1\leqslant G^{(i)} \leqslant3$.

\subsection{Caveats}
\label{Subsec-debate on the model}

We should mention several caveats regarding our model. The first one is the hypothesis of a thin disk, that helps us to simplify the equations and the calculations. Real disks (from simulations \citep{Bate_2018,Lebreuilly_ea_2021} and observations \citep{Andrews_ea_2018a,Andrews_ea_2018b,Sheehan_ea_2022a,Sheehan_ea_2022b,Tobin_ea_2020a,Tobin_ea_2020b}) look more fluffy and do not have a perfect shape. These differences in shape and sizes certainly leads to different physics and more complicated equations (in real case). But our intention is to regard only the mid-plane of a fluffy disk where the more material is concentrated, and hence this thin volume dominates the physics of the whole disk, at least in the sense of energies per unit mass, what is a key point for our approach.

The second one is the neglecting of the magnetic fields, which are found to have a huge influence on the size, shape and dynamics of the fluid in disk \citep{Lebreuilly_ea_2021,Hennebelle_ea_2022}. Also, as one can see in the work by \cite{Mocz_ea_2017} the magnetic fields play significant role even at larger scales. In the latter work the authors present simulations of the collapse of prestellar cores in supersonic, turbulent, isothermal, magnetized environments, exploring the effect of the mean magnetic field strength. The density profile is $\varrho\propto\ell^{-2}$, which means that the scales in regard correspond to the outer shells in respect to the protostellar disk (which is not formed yet). This can be seen in their fig.5, where they present the scaling of the gas pressure, magnetic pressure, kinetic energy density, and gravitational potential energy density. Although, in their fig.4 one can see that the density PDFs are getting shallower for stronger B-field simulations. Hence we expect that magnetic fields will have even greater impact at protostellar disk's scales. Their consideration needs, however, much more complicated physical model and hence more and more involved mathematics. So we regard this as a next step in the building of more reliable model. For now we are satisfied with a simpler hydrodynamical model.

The third one is the neglecting of a feedback from new-born stars in the core. This is related to magnetic fields, out-flows and radiation that new-born stars cause. Of course, the latter factors strongly affect the shape and size of the disk, also its temperature, and the accretion rate of material (see the simulation of \cite{Lebreuilly_ea_2023}). This considerable simplification we regard as a necessary step to simplify our first approximation model.

The fourth one is the disregard of the viscosity in the disk, which is proved to explain the disk evolution at its late stages (see \cite{Lynden-Bell_Pringle_1974}), when the protostar is always burning. The latter simplification we justify with the assumption that our model regards the system at early stages after its formation. Also, if the viscosity scales according to the power-law $\ell^{\gamma}$, where $\gamma\geqslant1$ (see \cite{Lynden-Bell_Pringle_1974} and \cite{Hartmann_ea_1998}), then we expect it will be vanishing for $\ell\ll1$, where our solutions are valid.

The fifth is the assumption for steady state. As it shown in \cite{Hartmann_ea_1998} the accretion rate of the material at the disk boundary is not a constant with time, but rather decline during the disk evolution. So the assumption for steady state will be broken. But if one takes into account that the changes of the large-scale velocity field (the accretion) are slow, and the establishing of local pressure equilibrium (through the sound speed), near to the core, is much faster, then comparing the typical time-scales of these processes one may conclude that the steady state hypothesis can be substantiated. We suppose that there exists a period of time when the cloud reaches a nearly steady state \citep{Burkert_2017} and its PDF is characterized by two PLTs in the high density range, with roughly constant slopes \citep{KNW_2011,Girichidis_ea_2014,Schneider_ea_2015c}. Our model reproduces those two slopes in approximate solutions (Paper I, Paper II, \cite{Donkov_ea_2021,Donkov_ea_2022}, and this work).

We are aware of the considered model is highly idealized and also not applicable to many real objects. But, indeed, our goal is not to restore the complete dynamics and morphology of the fluid near to the core, but rather to assess in first approximation its general characteristics in purely hydrodynamical case.

\section{Conclusions}
\label{Sec-conclusions}

In the current paper an equation for the density profile (see equation (\ref{Ealg_rho})) of a self-gravitating polytropic turbulent disk, accreting material from its vicinity and immersed in a molecular cloud is obtained. The derived profile describes the conditions of the medium near to the cloud core, where a protostar/protocluster is forming appropriately. The model is based on the framework established in \citet{DS_2018, DS_2019}. Unlike these papers, here the EOS of the gas near to the very small and dense core has been changed from isothermal ($\Gamma=1$) to polytropic ($\Gamma>1$), like in \citet{Donkov_ea_2021}, but compared to the latter paper the symmetry in the core vicinity is not spherical but rather cylindrical, namely rotating disk. In order to prove the invariance of the total energy per unit mass with respect to the fluid flow we assume that the accretion of matter through the boundary of the cloud and through all spatial scales down to the core is quasi-static, i.e. the cloud is in steady state. The equation that we obtain for the density profile function, or the generalized Bernoulli equation of our model, which describes the balance of the different types of energies per unit mass of a fluid element -- the kinetic, the thermal, and the gravitational, is a non-linear integral one (see equations (\ref{dE/dl=0_rho}), (\ref{E_rho}) and (\ref{Ealg_rho})).
	
The method that we apply for the obtaining of an approximate solution for the density profile of the form $\rho(l)\propto l^{-p}$ allows us to evaluate the power-law PDF which describes the dense cloud regions, where self-gravity cannot be neglected. The PDF slope is consistent with estimates from numerical simulations \citep{KNW_2011, Veltchev_ea_2019,Marinkova_ea_2021} and some observations \citep{Schneider_ea_2015c,Schneider_ea_2022}. 

Our results show three regimes characterized by different density profiles, polytropic exponents, angular velocity scaling exponents and energy balance equations for the fluid elements. Which one of them occurs depends of the balance of the centrifugal kinetic term and the thermal term, on the one side, and the gravitational term, on the other. Then applying to the model dynamical analysis (see equation (\ref{equ-force balance})) and comparing the result from it to the regimes, obtained through the energy equation, one might conclude that the only reasonable approximate solution is given by the second regime, where the gas pressure force balances the gravity and the rotation of the disk is sub-Keplerian. Accordingly, the energy balance equation for a fluid element reads: $ p\Gamma - G^{\rm (i)} \approx 0$, and the polytropic exponent is given by the formula: $\Gamma = 1+1/p$, hence $G^{(i)}=1+p$. Taking into account that we suppose the density profile exponent $p$ to be in the range $1\leq p \leq2$, this gives the ranges for $\Gamma$ and $G^{(i)}$. They are $2\geq \Gamma \geq 3/2$ and $2\leq G^{(i)} \leq3$, respectively. The above-mentioned second physical regime applies to the detection of such power-law tail in (column-)density PDFs in some observational and numerical studies of dense cloud regions \citep{Schneider_ea_2015c,Schneider_ea_2022, KNW_2011, Marinkova_ea_2021}. The extracted first power-law tail in PDFs at lower densities can be assessed from the consideration of the isothermal outer shells \citep{DS_2018, DS_2019,Donkov_ea_2022}.

At the end we would like to state that we are aware of the constraints of our results for the second power-law tail in the density PDF, which determines the structure of the inner regions of molecular clouds. In our consideration we have neglected factors such as magnetic fields, viscosity and feedback from new-born stars, which may be significant. A substantiated explanation of the second power-law tail should come from a more complex model and our work can be considered as a contribution to such efforts.

ACKNOWLEDGEMENT
The authors are indepted to the reviewer for the valuable comments and recommendations, and especially for the proposal for inclusion of a table which summarizes our main results. S.D. acknowledges funding by the Ministry of Education and Science of Bulgaria (support for the Bulgarian National Roadmap for Research Infrastructure) and for the support by DFG under grant KL 1358/20-3. T.V. acknowledges support by the DFG under grant KL 1358/20-3 and additional funding from the Ministry of Education and Science of the Republic of Bulgaria, National RI Roadmap Project DO1-176/29.07.2022. R.S.K. acknowledges financial support from the German Research Foundation (DFG) via the Collaborative Research Center (SFB 881) 'The Milky Way System' (sub-projects: B1, B2, and B8) and in the Priority Program SPP 1573 "Physics of the Interstellar Medium" (grant numbers KL 1358/18.1, KL 1358/19.2).

DATA AVAILABILITY

No new data were generated or analysed in support of this research.

\label{lastpage}

\end{document}